\documentstyle[12pt]{article}
\begin{document}
\title{ Thermal and Nonthermal Pion Enhancements with Chiral Symmetry Restoration }
\author{Pengfei Zhuang and Zhenwei Yang\\
        Physics Department, Tsinghua University, Beijing 100084, China}
\date{}
\maketitle

\begin{abstract}
The pion production by sigma decay and its relation with chiral symmetry restoration in a hot and dense matter are investigated in the framework of the Nambu-Jona-Lasinio model. The decay rate for the process $\sigma\rightarrow 2\pi$ to the lowest order in a $1/N_c$ expansion is calculated as a function of temperature $T$ and chemical potential $\mu$. The thermal and nonthermal enhancements of pions generated by the decay before and after the freeze-out are present only in the crossover region of the chiral symmetry transition. The strongest nonthermal enhancement is located in the vicinity of the endpoint of the first-order transition.  
\end{abstract}

\noindent ${\bf PACS: 05.70.Jk, 11.30.Rd, 14.40.-n, 25.75.-q}$

\section {Introduction}

It is generally believed that there are two QCD phase transitions in hot and dense nuclear matter\cite{muhe}. One of them is related to the deconfinement process in moving from a hadron gas to a quark-gluon plasma, and the other one describes the transition from the chiral symmetry breaking phase to the phase in which it is restored. Since the deconfinement and restored phases are only intermediate states in relativistic heavy ion collisions which are considered to be the way to generate the two transitions in laboratories, the key problem is how to extract possible signatures of the phase transitions hidden in the final state distributions. 

One way to study the signatures of the chiral transition is to compare the properties of hadronic systems with and without undergoing a chirally symmetric phase. In recent years many such signatures have been proposed, such as the excess of pions due to a rapid thermal and chemical equilibration in the symmetric phase\cite{excess1}, the excess of low energy photon pairs by pion annihilation in hot and dense medium\cite{excess2}, the dilepton enhancement from leptonic decay of sigma\cite{dilepton}, and the enhancement of the continuum threshold in the scalar channel\cite{continum}. Most of the signatures are associated with the changes of sigma properties at finite temperatures and densities. In the vacuum there is no obvious reason for the presence of the scalar meson sigma in the study of chiral properties. However, there is a definite need\cite{rho} for sigma at finite temperatures and densities in the investigation of chiral symmetry restoration. With increasing $T$ and $\mu$ sigma changes its character from a resonance with large mass to a bound state with small mass. This medium dependence of sigma mass makes important sense not only in the above mentioned particle yields, but also in thermodynamics\cite{zhuang}. The contribution from sigma to any thermodynamic function  can for most case be neglected, because it is heavy, while it plays an important role in the region around the critical point of the chiral phase transition, since there $m_\sigma=0$. 

From the recent QCD model calculations\cite{tric1}, the chiral phase transition in the chiral limit with massless pions is of second order at high temperatures, and of first order at high densities. Therefore, there is a tricritical point $P$ in the $T-\mu$ plane which separates the first- and second-order phase transition. In the real world with nonzero pion mass, the second order phase transition becomes a smooth crossover and the tricritical point $P$ becomes an endpoint $E$ of a first-order phase transition line. In relativistic heavy ion collisions, for the choice of control parameters like colliding energy and impact parameter such that the freeze-out of the system occurs near the point $E$, sigma is one of the most numerous species at the freeze-out, since it is nearly massless. These light sigmas at the freeze-out will lead to large event-by-event fluctuations\cite{tric2} of final state pions due to the $\sigma$-exchange in the pion thermodynamic potential. Another direct signature of this critical endpoint $E$ is a nonthermal excess\cite{tric2} of pions from the hadronic decay of sigma. With the expansion of the hadronic system produced in relativistic heavy ion collisions, the in-medium sigma mass rises towards its vacuum value and eventually exceeds the $\pi\pi$ threshold. As the $\sigma\pi\pi$ coupling is large, the decay proceeds rapidly. Since this process occurs after freeze-out, the pions generated by it do not have a chance to thermalize. Thus, it is expected\cite{tric2} that the resulting pion spectrum will have a nonthermal enhancement at low transverse momentum. 

The present paper is devoted to a systematic and detailed investigation of the above idea\cite{tric2} on excess pions produced by sigma decay in a chiral quark model in the whole $T-\mu$ plane, without the assumption that the freeze-out occurs at the chiral transition. In the chiral limit, we know very well that the sigma decay into two pions starts at the phase transition point, and the freeze-out happens later. Therefore, the sigma decay before and after the freeze-out will result in thermal and nonthermal enhancements of pions, respectively. Our motivation here is to see that in which temperature and density region of freeze-out there is a strong nonthermal enhancement. 

We proceed as follows. In Section 2 we calculate the sigma decay rate as a function of $T$ and $\mu$ to the lowest order in a $1/N_c$ expansion in the framework of the Nambu-Jona-Lasinio model (NJL) at quark level\cite{njl}, and discuss the importance of the Bose-Einstein statistics at high $T$ and the dominance of the first-order transition at high $\mu$. The in-medium pion mass drops down with decreasing $T$ and $\mu$. When it approaches to its vacuum value, the freeze-out happens. With this definition of freeze-out and the help of scaling hydrodynamics to describe the expansion of the hadronic system, we study in Section 3 the thermal and nonthermal pion enhancements due to $\sigma$ decay. Finally we give our summary and point out open problems that remain.

\section { sigma decay rate at finite temperatures and densities }

One of the models that enables us to see directly how the dynamical mechanisms of chiral symmetry breaking and restoration operate is the NJL model applied to quarks\cite{njl}. Within this model one can obtain the hadronic mass spectrum and the static properties of mesons remarkably well. In particular, one can recover the Goldstone mode, and some important low-energy properties of current algebra such as the Goldberger-Treiman and GellMann-Oakes-Renner relations. Because of the contact interactions between quarks that is introduced in this model, there is of course no confinement. A further consequence of this feature is that the model is non-renormalizable, and it is necessary to introduce a regulator $\Lambda$ that serves as a length scale in the problem. The sigma decay rate at zero density in this model was firstly studied by Hatsuda and Kunihiro\cite{hatsuda}. In order to investigate the thermal and nonthermal pion yields by sigma decay, we recalculate here the decay rate in the whole $T-\mu$ plane, and analyze its relation with the first-order transition in high density region.  

The two-flavor version of the NJL model is defined through the lagrangian density,
\begin{equation}
\label{njl}
L_{NJL} = \bar\psi (i\gamma^\mu\partial_\mu - m_0)\psi+G[(\bar\psi\psi)^2+(\bar\psi i \gamma_5\tau\psi)^2],
\end{equation}
where only the scalar and pseudoscalar interactions corresponding to $\sigma$ and $\pi$ mesons, respectively, are considered, $\psi$ and $\bar\psi$ are the quark fields, the operate ${\bf \tau}$ is the $SU(2)$ isospin generator, $G$ is the coupling constant with dimension $GeV^{-2}$, and $m_0$ is the current quark mass.

In an expansion in the inverse number of colors, $1/N_c$, the zeroth order (Hartree) approximation for quarks together with the first-order (RPA) approximation for mesons gives a self-consistent treatment of the quark-meson plasma in the NJL model. In the chiral limit the tricritical point $P$ is located at $T_P = 79\ MeV, \mu_P = 280\ MeV$ in the $T-\mu$ plane. In the case of explicit chiral symmetry breaking the endpoint of the first-order transition line is shifted with respect to the tricritical point $P$ towards large $\mu$, its position is $T_E = 23\ MeV, \mu_E = 325\ MeV$. 
To the lowest order, namely $\left(1/N_c\right)^{1/2}$, the Feynman diagram for the decay process $\sigma\rightarrow 2\pi$ is sketched in Fig.1. The partial decay rate in the rest frame of sigma is given in terms of the Lorentz-invariant matrix element $M$ by 
\begin{equation}
\label{rate1}
{d\Gamma_{\sigma\rightarrow 2\pi}\over d\Omega} = {1\over 32\pi^2}{|{\bf p}|\over m_\sigma^2}|M|^2\ ,
\end{equation}
where $|{\bf p}|=\sqrt{{m_\sigma^2\over 4}-m_\pi^2}$ is the pion momentum, $m_\sigma$ and $m_\pi$ are respectively $\sigma$- and $\pi$-mass. Using an obvious notation for the $M$-matrix element, one has
\begin{equation}
\label{matrix}
M=g_\sigma g_\pi^2 A_{\sigma\pi\pi}
\end{equation}
with
\begin{equation}
\label{aspp}
A_{\sigma\pi\pi} = T\sum_n e^{i\omega_n\eta}\int{d^3{\bf q}\over (2\pi)^3}Tr S({\bf q},i\omega_n)\Gamma_\pi S({\bf q}+{\bf p},i\omega_n+{m_\sigma\over 2})\Gamma_\pi S({\bf q},i\omega_n +m_\sigma)\ .
\end{equation}
Here $\omega_n = (2n+1)\pi T, n=0,\pm 1,\pm 2,...$ are fermionic Matsubara frequencies, $\Gamma_\pi = i\gamma_5\tau$ is the pseudoscalar interaction vertex. The quark propagator is denoted by $S({\bf q},i\omega_n) = \left(\gamma_0(i\omega_n+\mu)-{\bf \gamma}\cdot{\bf q}+m_q\right)/\left((i\omega_n+\mu)^2-E_q^2\right)$ with the quark energy $E_q = \sqrt{m_q^2+{\bf q}^2}$. In Eq.(\ref{aspp}), $Tr$ refers to the trace over color, flavor and spinor indices. The pion-quark-quark and sigma-quark-quark coupling strengths in Eq.(\ref{matrix}) are determined in the model via\cite{njl}
\begin{eqnarray}
\label{strength}
&& g_\pi^{-2}(T,\mu) = {\partial \Pi_\pi(k_0,{\bf 0};T,\mu)\over \partial k_0^2}|_{k_0^2=m_\pi^2}\ ,\nonumber\\
&& g_\sigma^{-2}(T,\mu) = {\partial \Pi_\sigma(k_0,{\bf 0};T,\mu)\over \partial k_0^2}|_{k_0^2=m_\sigma^2}\ ,
\end{eqnarray}
where $\Pi_\pi$ and $\Pi_\sigma$ are the standard mesonic polarization functions\cite{njl} for pion and sigma. The dynamically generated quark mass $m_q(T,\mu)$ and the meson masses $m_\pi(T,\mu)$ and $m_\sigma(T,\mu)$ are calculated using the usual gap equations\cite{njl} in the Hartree approximation for quarks and RPA approximation for mesons. After evaluation of the Matsubara sum in Eq.(\ref{aspp}), one has\cite{hufner}
\begin{eqnarray}
\label{aspp1}
A_{\sigma\pi\pi}(T,\mu) = && 4m_q N_c N_f\int{d^3{\bf q}\over (2\pi)^3}{f_F(E_q-\mu)-f_F(-E_q+\mu)\over 2E_q}\times \nonumber\\
&& {8({\bf q}\cdot{\bf p})^2-(2m_\sigma^2+4m_\pi^2){\bf q}\cdot{\bf p}+m_\sigma^4/2-2m_\sigma^2 E_q^2\over (m_\sigma^2-4E_q^2)\left((m_\pi^2-2{\bf q}\cdot{\bf p})^2-m_\sigma^2 E_q^2\right)}\ ,
\end{eqnarray} 
where $f_F(x) = \left(1+e^{x/T}\right)^{-1}$ is the Fermi-Dirac distribution function.

Sigma can decay into neutral and charged pions. By considering the exchange contribution for $\sigma\rightarrow 2\pi_0$ due to the identical particle effect, the total decay rate can be written as
\begin{eqnarray}
\label{rate2}
\Gamma_{\sigma\rightarrow 2\pi}(T,\mu) && = \Gamma_{\sigma\rightarrow 2\pi_0}(T,\mu)+\Gamma_{\sigma\rightarrow \pi_+\pi_-}(T,\mu)\nonumber\\
&& = {3\over 8\pi}{\sqrt{{m_\sigma^2\over 4}-m_\pi^2}\over m_\sigma^2}g_\sigma^2 g_\pi^4 |A_{\sigma\pi\pi}(T,\mu)|^2\left(1+2f_B({m_\sigma\over 2})\right)\ ,
\end{eqnarray}
where we have taken into account the Bose-Einstein statistics in terms of the distribution function $f_B(x) = \left(e^{x/T}-1\right)^{-1}$ for the final state pions\cite{hatsuda}. 
    
From the comparison of $\sigma\pi\pi$ coupling strength $g_{\sigma\pi\pi}$ defined through the lagrangian $L_I \sim g_{\sigma\pi\pi}\sigma{\bf \pi}{\bf \pi}$ and the $M$-matrix element (\ref{matrix}) in the NJL model, we have
\begin{equation}
\label{gspp}
g_{\sigma\pi\pi}(T,\mu) = 2M(T,\mu) = 2g_\sigma g_\pi^2 A(T,\mu)\ .
\end{equation}

In the numerical calculations we used the dynamical quark mass $m_q = 0.32\ GeV$, the pion decay constant $f_\pi = 0.093\ GeV$ and the pion mass $m_\pi = 0.134\ GeV$ in the vacuum ($T=\mu=0$) as input to determine the $3$ parameters in the NJL model, namely the coupling constant $G=4.93\ GeV^{-2}$, the momentum cutoff $\Lambda=0.653\ GeV$ and the current quark mass $m_0 = 0.005\ GeV$.

We first discuss the temperature dependence of $g_{\sigma\pi\pi}$ and $\Gamma_{\sigma\rightarrow 2\pi}$ in the case of zero baryon density which corresponds to the central region of relativistic heavy ion collisions. The coupling strength $g_{\sigma\pi\pi}$ is about $2\ GeV$ in the vacuum and almost a constant in the temperature region $T< 0.1\ GeV$, and then decreases smoothly with increasing temperature, as shown in Fig.2. The energy condition for the decay process $\sigma\rightarrow 2\pi$ to happen is $m_\sigma \geq 2m_\pi$ in the rest frame of $\sigma$. At the threshold temperature $T\simeq 0.19\ GeV$ determined by $ m_\sigma(T)=2m_\pi(T)$, $g_{\sigma\pi\pi}$ has a sudden jump from about $0.3 /fm$ to zero. In a wide temperature region, the decay rate (\ref{rate2}) goes up gradually from its vacuum value $\sim 0.45/fm$ to the maximum $\sim 0.5/fm$ at $T\simeq 0.17\ GeV$, then it drops down rapidly but continuously and becomes zero at the threshold temperature. The fact that the maximum decay rate is not in the vacuum where the mass difference between $\sigma$ and $\pi$ has the largest value, but at $T\simeq 0.17\ GeV$ is beyond our expectation. It originates from the important Bose-Einstein statistics of the final state pions at high temperatures. This can be seen clearly in Fig.2 where we show the decay rates with and without consideration of the Bose-Einstein statistics. The rapid decrease of $\Gamma_{\sigma\rightarrow 2\pi}$ at high temperatures is due to the behavior of $g_{\sigma\pi\pi}$ and the phase space factor $\sqrt{m_\sigma^2/4-m_\pi^2}/m_\sigma^2$.    

The density dependence of $g_{\sigma\pi\pi}$ and $\Gamma_{\sigma\rightarrow 2\pi}$ is shown in Fig.3. Since the threshold temperature for $\sigma$ decay decreases with increasing chemical potential, the crucial Bose-Einstein statistics effect at high temperatures is gradually washed out as the chemical potential goes up. There is almost no difference between with and without such statistics when $\mu \ge 0.25$. Instead of the statistics the first-order chiral transition dominates the behavior of $g_{\sigma\pi\pi}$ and $\Gamma_{\sigma\rightarrow 2\pi}$ at high densities. When the system passes through the first-order transition line from the chirally symmetric phase to the broken phase, all the masses, $m_q, m_\pi$ and $m_\sigma$ are discontinuous, $m_q$ and $m_\sigma$ jump up and $m_\pi$ jumps down. Therefore, the mass 
difference between $\sigma$ and $\pi$ has a jump from less than $2m_\pi$ to larger than $2m_\pi$, and the $\sigma$ decay starts not at the threshold, but at some value beyond the threshold. This means that the decay in the first-order transition region happens suddenly with a nonzero rate, the temperature dependence of the decay rate behaves like a step function. This is shown in Fig.3 for $\mu = 0.33 GeV \ge \mu_E$. When approaching to the endpoint $E$ from the first-order transition side, the jump of the mass difference disappear. The decay rate becomes continuous, while the coupling strength has still a jump.

To see the parameter dependence of the above discussion, we considered another set of parameters, $G=5.46\ GeV^{-2}, \Lambda=0.632\ GeV$ and $m_0=0.0055\ GeV$. It makes shifts of the magnitudes of $g_{\sigma\pi\pi}$ and $\Gamma_{\sigma\rightarrow 2\pi}$, but does not change their shapes, especially the behavior around the threshold temperature or the chiral phase transition point. 

Let's summarize the calculations of the $\sigma$ decay rate. When the chemical potential is extremely high, the first-order chiral transition results in an almost temperature-independent decay rate. For most of the decay processes in the crossover region, the maximum decay rate does not locate in the vacuum, but is very close to the threshold point. Only when the chemical potential is around $0.25\ GeV$ which is close to the endpoint $E$, the decay rate increases continuously and not very fast in the beginning, and then it behaves like a constant. As we discussed in the introduction, the freeze-out of the pions may occur after $\sigma$ decay. According to the decay rate shown in Fig.3, when the decay happens in the first-order transition region or most of the places in the crossover region in the $T-\mu$ plane, a large fraction of $\sigma$'s may decay into pions before the freeze-out, and these pions are thus thermalized before their freeze-out, only those pions generated after the freeze-out contribute to the nonthermal enhancement. However, when the $\sigma$ decay happens with $\mu \sim 0.25 GeV$, most of the pions may be produced after the freeze-out, and therefore, there may be a strong nonthermal enhancement in this case. In the next section we discuss the nonthermal enhancement quantitatively.

\section { thermal and nonthermal pion enhancements }

Before we calculate the pion numbers with the $\sigma$ decay rate as a function of temperature $T$ and chemical potential $\mu$ given in the last section, we need to treat $T,\mu$ as time-dependent. In the baryon-free ($\mu = 0$) region of relativistic heavy ion collisions, the Bjorken's scaling hydrodynamics\cite{bjorken} is often used to describe the space-time evolution of hadronic thermodynamics. The time dependence of temperature in this scenario is given by 
\begin{equation}
\label{expansion}
T=T_d\left({t_d\over t}\right)^\alpha
\end{equation}
with $\alpha = 1/3$. Here we have defined $t_d$ to be the time at which $\sigma$ decay begins and the threshold temperature $T_d$ satisfies the mass relation $m_\sigma(T_d) = 2m_\pi(T_d)$. We will 
be interested in times $t\geq t_d$. In hydrodynamic models\cite{bjorken,kajantie} the initial time  corresponding to the highest temperature of the fluid is normally taken to be $\sim 1 fm$. Since the sigma decay starts in the early stage of the fluid, we choose $t_d = 2 fm$ in our numerical calculations.

For heavy ion collisions with full nuclear stopping, the lifetime of the baryon-rich fluid is much shorter compared with that of the baryon-free fluid, the space evolution can be considered as a homogeneous 3-dimensional expansion (\ref{expansion}) with $\alpha = 1$ instead of $\alpha = 
1/3$. The real relativistic heavy ion collisions are between these two limits. We still use the expansion mechanism(\ref{expansion}) as the time dependence of temperature for any chemical potential $\mu$, but the exponent $\alpha$ is now $\mu$-dependent and parameterized by
\begin{equation}
\label{alpha}
\alpha(\mu) = \left(1+2{\mu\over \mu_{max}}\right)\alpha(0)
\end{equation}
with $\alpha(0) = 1/3$. Here the maximum chemical potential $\mu_{max}$ corresponding to the full stopping limit is chosen as $0.4\ GeV$ in the NJL model.

With the known decay rate $\Gamma(T,\mu)$ and the relation between temperature and time for any chemical potential, the number of $\sigma$'s present at time $t$ is determined by
\begin{equation}
\label{sigma1}
{dN_\sigma (t)\over dt} = -\Gamma(t) N_\sigma(t)\ ,
\end{equation}
and is therefore
\begin{equation}
\label{sigma2}
N_\sigma (t) = N_\sigma (t_d) e^{-\int_{t_d}^t \Gamma(t')dt'}\ ,
\end{equation}
where $N_\sigma (t_d)$ is the maximum number at the beginning time $t_d$,
\begin{eqnarray}
\label{sigma3}
&& N_\sigma (t_d) = V(t_d)n_\sigma(t_d)\ ,\nonumber\\
&& n_\sigma (t_d) = \int {d^3{\bf p}\over (2\pi)^3}{1\over e^{\epsilon_\sigma(T_d)/T_d}-1}
\end{eqnarray}
with the $\sigma$ energy $\epsilon_\sigma = \sqrt{m_\sigma^2+{\bf p}^2}$ and the volume $V(t_d)$ of the system at time $t_d$.

Since each sigma yields two pions, the number of pions generated by $\sigma$ decay at time $t$ is related to the number of sigma by
\begin{equation}
\label{pion}
N_\pi (t) = 2\left(N_\sigma (t_d) - N_\sigma (t)\right)\ .
\end{equation}
As the time $t \rightarrow \infty$, $N_\sigma (t \rightarrow \infty) = 0$, sigma disappears, and the pion number reaches its maximum $N_\pi (t\rightarrow \infty) = 2N_\sigma (t_d)$.

The conventional definition for freeze-out is that the time between two successful scattering events is larger than the collective expansion time scale\cite{heinz},
\begin{equation}
\label{freezeout}
\tau_{scatt} \ge \tau_{exp}\ .
\end{equation}
$\tau_{scatt}$ can be calculated\cite{zhk} from the thermally averaged $\pi\pi$ cross section\cite{qzk} in the NJL model, and $\tau_{exp}=1/\partial_\mu u^\mu$ is simply reduced to $\tau_{exp} = t$ in the scaling hydrodynamics\cite{bjorken}. Thus the freeze-out time $t_f$ and the freeze-out temperature $T_f$ are determined by $\tau_{scatt} = t_f$ and the expansion mechanism (\ref{expansion}). With the two temperature scales $T_d$ and $T_f$ we can divide the total pions generated by $\sigma$ decay into thermal and nonthermal pions. For $T_d > T_f$, the produced pions in the interval $T_f<T<T_d$ have time to thermalize before the freeze-out, they lead to an enhancement of thermal pions, while in the interval $T<T_f$ the produced pions do not get a chance to thermalize, they result in a nonthermal enhancement of pions with low momentum. For $T_d<T_f$, $\sigma$ decay yields no thermal pions, all the resulted pions are nonthermal pions. 

The thermal and nonthermal enhancements are defined as ratios of the numbers of the pions produced by $\sigma$ decay before and after the freeze-out to the number of the pions produced directly by the freeze-out of the original thermal pion gas,
\begin{eqnarray}
\label{thnt}
&& r_{th}(t) = {N_\pi(t)\over N_\pi^{dir}(t_f)}\ ,\ \ \ \ \ t_d\leq t\leq t_f\ ,\nonumber\\
&& r_{nt}(t) = {N_\pi(t)-N_\pi(t_f)\over N_\pi^{dir}(t_f)}\ ,\ \ \ \ \ t_f\leq t
\end{eqnarray}
where $N_\pi^{dir}(t_f)$ is determined by the freeze-out temperature $T_f$, the volume of the system at time $t_f$, and the pion energy $\epsilon_\pi = \sqrt{m_\pi^2+{\bf p}^2}$, 
\begin{eqnarray}
\label{dir}
&& N_\pi^{dir}(t_f) = V(t_f) n_\pi^{dir}(t_f)\ ,\nonumber\\ 
&& n_\pi^{dir}(t_f) = \int{d^3{\bf p}\over (2\pi)^3}{3\over e^{\epsilon_\pi(T_f)/T_f}-1}\ .
\end{eqnarray}
Obviously the measurable thermal and nonthermal enhancements are $r_{th}(t_f)$ and $r_{nt}(t\rightarrow\infty)$.

With the in-medium pion momentum $p=\sqrt{{m_\sigma^2\over 4}-m_\pi^2}$, the expansion mechanism (\ref{expansion}), and the pion distribution (\ref{pion}), we determine the normalized momentum distribution of the nonthermal pions,
\begin{equation}
\label{nmd}
{1\over N_{nt}}{dN_{nt}\over dp} = {e^{-\int_{t_d}^t \Gamma(t')dt'}\Gamma(t)\theta(t-t_f){dt\over dp}\over \int^\infty_{t_f}
e^{-\int_{t_d}^t \Gamma(t')dt'}\Gamma(t) dt}\ ,
\end{equation}
and the observable mean momentum of the nonthermal pions, 
\begin{equation}
\label{avep}
\bar p_{nt} = {\int p{dN_{nt}\over dp}dp \over \int {dN_{nt}\over dp}dp}        
= {\int_{t_f}^\infty p(t)e^{-\int_{t_d}^t \Gamma(t')dt'}\Gamma(t) dt\over \int_{t_f}^\infty e^{-\int_{t_d}^t \Gamma(t')dt'}\Gamma(t) dt}\ .
\end{equation}
Since it is difficult to separate out the nonthermal pions experimentally, we calculate also the normalized momentum spectra of the total pions including the direct pions, the thermal pions from sigma decay, and the nonthermal pions from sigma decay,
\begin{eqnarray}
\label{tmd}
&& {1\over N_{total}}{dN_{total}\over dp} = {1\over n_\pi^{dir}(t_f)+2{t_d\over         t_f}n_\sigma(t_d)}\times\\
&& \left(\left(1+2{t_d\over t_f}{n_\sigma(t_d)\over n_\pi^{dir}(t_f)}\left(1-
   e^{-\int_{t_d}^{t_f} \Gamma(t')dt'}\right)\right){dn_\pi^{dir}(t_f,p)\over dp}
   +2{t_d\over t_f}n_\sigma(t_d)e^{-\int_{t_d}^t \Gamma(t')dt'}\Gamma(t){dt\over dp}\theta(t-t_f)\right)\ ,\nonumber
\end{eqnarray}
where we have used
\begin{equation}
\label{volume}
{V(t_d)\over V(t_f)} = {t_d\over t_f}
\end{equation}
in the scaling hydrodynamics\cite{bjorken}, and ${dn_\pi^{dir}(t_f,p)\over dp}$ is defined as
\begin{equation}
\label{dndp}
{dn_\pi^{dir}(t_f,p)\over dp}={p^2\over 2\pi^2}{3\over e^{\epsilon_\pi(T_f)/T_f}-1}\ .
\end{equation}
 
We now discuss our numerical calculations about the thermal and nonthermal pion distributions. Fig.4 shows the sigma number and pion number scaled by the corresponding maximum value,
\begin{eqnarray}
\label{dis1}
&& r_\sigma(t) = {N_\sigma(t)\over N_\sigma(t_d)}\ ,\nonumber\\
&& r_\pi(t) = {N_\pi(t)\over N_\pi(t\rightarrow\infty)} = 1 - r_\sigma(t)\ ,
\end{eqnarray}
and the thermal and nonthermal enhancements $r_{th}(t)$ and $r_{nt}(t)$ as functions of time scaled by the beginning time $t_d$ in the case of $\mu = 0.2 GeV$. The dashed line located at the freeze-out time $t_f$ separates the left thermal region and the right nonthermal region. $48\%$ of the $\sigma$'s decay before the freeze-out, the rest decay into nonthermal pions after the freeze-out. The measurable thermal and nonthermal enhancements compared with the number of direct pions are $32\%$ and $30\%$, respectively.

The measurable thermal, nonthermal and total pion enhancements 
\begin{eqnarray}
\label{menhance}
&& r_{th}(t_f) = 2 {N_\sigma(t_d)\over N_\pi^{dir}(t_f)}(1-r_\sigma(t_f))\ ,\nonumber\\
&& r_{nt}(t\rightarrow \infty)=2{N_\sigma(t_d)\over N_\pi^{dir}(t_f)}r_\sigma(t_f)\ ,\nonumber\\
&& r_{th}(t_f)+r_{nt}(t\rightarrow \infty) = 2{N_\sigma(t_d)\over N_\pi^{dir}(t_f)}
\end{eqnarray}
are shown in Fig. 5 as functions of chemical potential $\mu$. At a first-order chiral transition point any particle undergoes a mass jump. At the low end of this jump, $m_\sigma < 2m_\pi$, $\sigma$ decay is forbidden, while at the high end, $m_\sigma > 2m_\pi$, the decay begins with a nonzero rate. This sudden decay was shown in Fig.3. On the other hand, the large $\sigma$ mass at the beginning of decay means a strong suppression of the sigma number $N_\sigma(t_d)$ which dominates the amplitudes of the enhancements, it drops down exponentially with increasing 
$\sigma$-mass. Compared with the decay happen in the crossover region where the decay begins with $m_\sigma = 2m_\pi$, the thermal and nonthermal pion enhancements are both washed away by the first-order chiral transition. This is clearly displayed in Fig.5. The enhancements start at the endpoint $E$ of the first-order transition line and develop in the crossover region. In a wide chemical potential interval, the total pion enhancement is very strong, $r_{th}(t_f)+r_{nt}(t\rightarrow\infty)\sim 0.5$. The shapes of the enhancements are mainly controlled by the decay rate shown in Fig.3. The nonthermal and total enhancements have their maximum values in the vicinity of the endpoint $E$, this is in agreement with our qualitative analysis in the last section.    

The momentum distribution of the nonthermal pions is shown in Fig.6 for four different chemical potentials. The nonthermal pions are always centralized in a very narrow momentum region around $p\simeq 0.26 GeV$. This narrow distribution leads to a sharp peak in the momentum spectra for the total pions, shown in Fig.7. For low chemical potentials the nonthermal momenta and the most appropriate thermal momenta almost coincide, while for high chemical potentials the nonthermal peak is separated from the thermal peak obviously. The thermal peak is always smooth, but the nonthermal peak is always sharp.

The mean momentum of the nonthermal pions is shown in Fig.8 as a function of chemical potential. Unlike the strong chemical potential dependence of the enhancements, especially around the endpoint $E$, the mean momentum is almost a constant, $\bar p_{nt} \sim 0.26 GeV$. This value is not so small as the people expected before\cite{tric2}. 

Our study on $\sigma$ decay depends on both the chiral dynamics which determines the decay rate, and the expansion mechanism which is reflected in the time evolution of the temperature. To see which one is dominant in the above calculations, we have chosen different expansion parameters. We changed $\mu_{max}$ to control the chemical potential dependence of the expansion rate, and the beginning time $t_d$ of the decay. While the amplitudes of the measurable thermal and nonthermal enhancements change remarkably, their shapes, the total enhancement and the mean momentum are insensitive to the expansion parameters. The chiral symmetry restoration plays a crucial rule in these observable quantities. We even let the freeze-out temperature $T_f$ be the decay temperature $T_d$, this means no thermal pions produced by $\sigma$ decay, the mean momentum in this case decreases by less than $10\%$ only.      
             
\section {conclusions}

Since the $\sigma$-mass goes up and the $\pi$-mass drops down as the hadronic system cools down, the more and more large mass difference between $\sigma$ and $\pi$ leads to $\sigma$ decay into two pions when the $\sigma$-mass exceeds the threshold $2m_\pi$. These pions result in thermal enhancement when they are produced before the freeze-out and nonthermal enhancement when they are produced after the freeze-out. Based on the two-flavor NJL model we have presented a detailed analysis of the pion production at finite temperatures and densities, and discussed the relation between the decay and the chiral symmetry restoration, especially the crucial effect of the order of the chiral transition on the thermal and nonthermal enhancements.

We have found that the chiral symmetry breaking and restoration control the pion enhancements of the system. At high densities the number of $\sigma$'s at the beginning of their decay is strongly suppressed by the large $\sigma$-mass, originating from the first-order chiral transition. Compared with the large number of direct pions produced by the freeze-out of the thermal pion gas, there are almost neither thermal nor nonthermal pion enhancements. In practice, the enhancements begin at the endpoint $E$ of the first-order chiral transition line and present only in the crossover region with high temperatures and low densities. The nonthermal enhancement in this region develops very fast in the early stage of the decay and reaches its maximum value in the vicinity of the endpoint $E$. For a wide chemical potential interval, $0\leq \mu \leq 0.25\ GeV$, both the thermal and nonthermal enhancements are remarkably large due to the almost constant decay rate during the whole decay process. The maximum enhancement is $63\%$ which agrees well with the value $2/3$ estimated by simple statistics\cite{tric2}.

While the total enhancement and the structure of the thermal and nonthermal enhancements seem insensitive to the expansion parameters we have chosen in the calculations, the amplitudes of the thermal and nonthermal enhancements depend strongly on the expansion mechanism. On the other hand, our results are derived in a particular chiral model. Although we expect the characteristics of the enhancements to be of more general validity, the numerical results are of course model dependent. Therefore, more reliable conclusions need more realistic expansion mechanism and more serious QCD model.

{\bf acknowledgments}:
This work was supported in part by the NSFC under grant numbers 19845001 and 19925519, and the Major State Basic Research Development Program under contract number G2000077407.

\newpage
{\bf Figure Captions}\\ \\ 

{\bf Fig.1}: 
The Feynman diagram for $\sigma$ decay into $2\pi$ to the lowest order in $1/N_c$ expansion. The solid lines denote quarks and antiquarks, and the dashed lines denote mesons.\\

{\bf Fig.2}:
The decay rate $\Gamma_{\sigma\rightarrow 2\pi}$ and the coupling strength $g_{\sigma\pi\pi}$ as functions of $T$ at $\mu = 0$. The thick and thin solid lines indicate the decay rates with and without consideration of Bose-Einstein statistics for the final state pions. \\

{\bf Fig.3}:
The decay rate $\Gamma_{\sigma\rightarrow 2\pi}$ as a function of $T$ for different chemical potentials. The lines marked with $a,b,c,d,e$ correspond to $\mu = 0,\ 0.1,\ 0.2,\ 0.25,\ 0.3\ GeV < \mu_E = 0.325\ GeV$, respectively, and the line $f$ is with $\mu = 0.33\ GeV > \mu_E$.\\

{\bf Fig.4}:
The sigma number and pion number scaled by the corresponding maximum value, and the thermal and nonthermal pion enhancements as functions of time scaled by the beginning time of the decay, at $\mu = 0.2 GeV$. The dashed line separates the left thermal and right nonthermal regions. \\

{\bf Fig.5}:
The measurable thermal, nonthermal and total pion enhancements as functions of $\mu$. \\

{\bf Fig.6}:
The normalized momentum distribution of the nonthermal pions for four different chemical potentials. \\

{\bf Fig.7}:
The normalized momentum spectra of the total pions for two different chemical potentials. \\

{\bf Fig.8}:
The mean momentum of the nonthermal pions as a function of $\mu$.\\


\newpage       
\begin{thebibliography}{99}
\bibitem{muhe}    {B.M\"uller, 
                   nucl-th/9906029;\\  
                   U.Heinz, 
                   hep-ph/9902424.}

\bibitem{excess1} {Chungsik Song, Volker Koch,  
                   Phys. Lett. {\bf B404}(1997)1.}

\bibitem{excess2} {M.K.Volkov, E.A.Kuraev, D.Blaschke, G.R\"opke, S.Schmidt,\\
                   Phys. Lett. {\bf B424}(1998)235.}

\bibitem{dilepton}{H.Arthur Weldon, 
                   Phys. Lett. {\bf B274}(1992)133.}

\bibitem{continum}{S.Chiku and T.Hatsuda,
                   Phys. Rev. {\bf D57}(1998)R6.}

\bibitem{rho}     {M.Rho, 
                   Proceedings of the International Workshop XXIII on Gross properties 
                   of Nuclei and Nuclear Excitations, Hirschegg, Austria, January, 16-21, 
                   1995.}

\bibitem{zhuang}  {P.Zhuang, J.H\"ufner, and S.P.Klevansky, 
                   Nucl. Phys. {\bf A576}(1994)525.}

\bibitem{tric1}   {R.Rapp, T.Sch\"afer, E.Shuryak and M.Velkovsky, 
                   Phys. Rev. Lett.{\bf 81}(1998)53;\\
                   M.Stephanov, K.Rajagopal and E.Shuryak,
                   Phys. Rev. Lett. {\bf 81}(1998)4816;\\
                   K.Rajagopal,
                   hep-ph/9808348;\\
                   J.Berges and K.Rajagopal, Nucl. Phys. {\bf B538}(1999)215.}

\bibitem{tric2}   {M.Stephanov, K.Rajagopal and E.Shuryak,
                   Phys. Rev. {\bf D60}(1999)114028.}

\bibitem{njl}     {For reviews and general references, see\\ 
                   U.Vogl and W.Weise,
                   Prog. Part. Nucl. Phys. {\bf 27}(1991)195;\\
                   S.P.Klevansky, 
                   Rev. Mod. Phys. {\bf 64}(1992)649;\\
                   M.K.Volkov, 
                   Prog. Part. Nucl. Phys. {\bf 24}(1993)35;\\
                   T.Hatsuda and T.Kunihiro, 
                   Phys. Rep. {\bf 247}(1994)338.}

\bibitem{hatsuda} {T.Hatsuda and T.Kunihiro,
                   Prog. Theor. Phys. Suppl. {\bf 91}(1987)284;
                   Phys. Lett. {\bf B185}(1987)304.}

\bibitem{hufner}  {J.H\"ufner, S.P.Klevansky, E.Quack and P.Zhuang, 
                   Phys. Lett. {\bf B337}(1994)30.}

\bibitem{bjorken}  {J.D.Bjorken,
                    Phys. Rev. {\bf D27}(1983)140.}

\bibitem{kajantie} {K.Kajantie, R.Ratio, P.V.Ruuskanen,
                    Nucl. Phys. {\bf B222}(1983)152;\\
                    Zhuang Pengfei, Wang Zhengqing and Liu Lianshou,
                    Z. Phys. {\bf C32}(1986)93;\\
                    Zhuang Pengfei and Liu Lianshou,
                    Z. Phys. {\bf C46}(1990)335.}

\bibitem{heinz}    {E.Schnedermann, J.Sollfrank, U.Heinz, in "Partical production in highly excited matter 
                    (eds. H.H.Gutbrod, J.Rafelski), NATO ASI Series B {\bf 303}(1993) 175 (Plenum, New York).}

\bibitem{zhk}      {P.Zhuang, J.H\"ufner, S.P.Klevansky and L.Neise, 
                    Phys. Rev. {\bf D51}(1995)3728.}

\bibitem{qzk}      {E.Quack, P.Zhuang, Yu.Kalinovsky, S.P.Klevansky and J.H\"ufner, 
                    Phys. Lett. {\bf B348}(1995)1;
                    M.Huang, P.Zhuanf and W.Chao,
                    Phys. Lett. {\bf B465}(1999)55.}
\end{thebibliography}
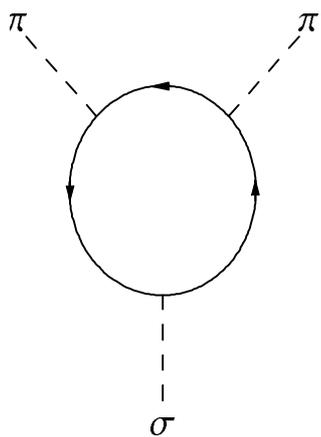
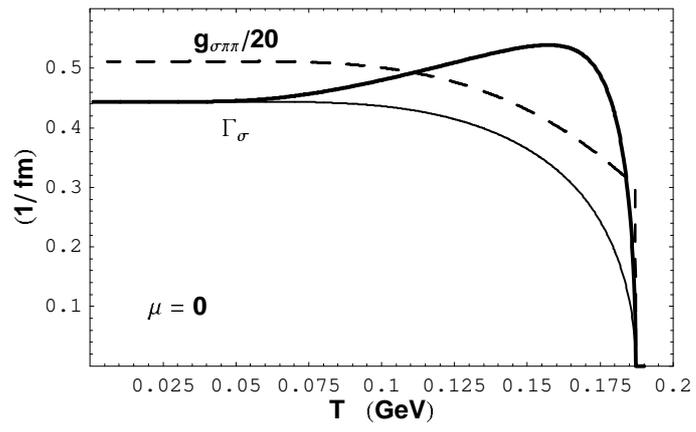
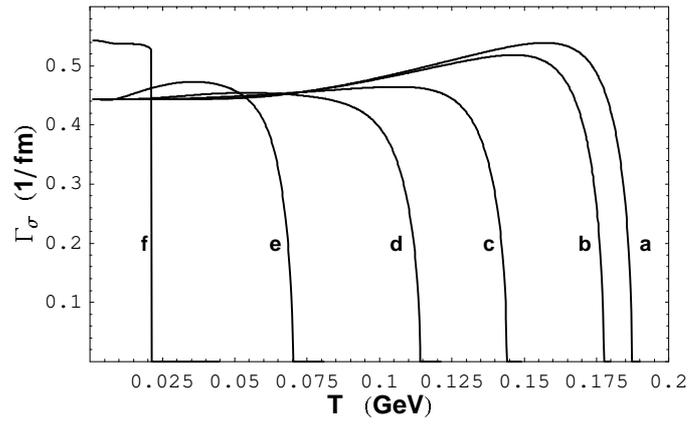
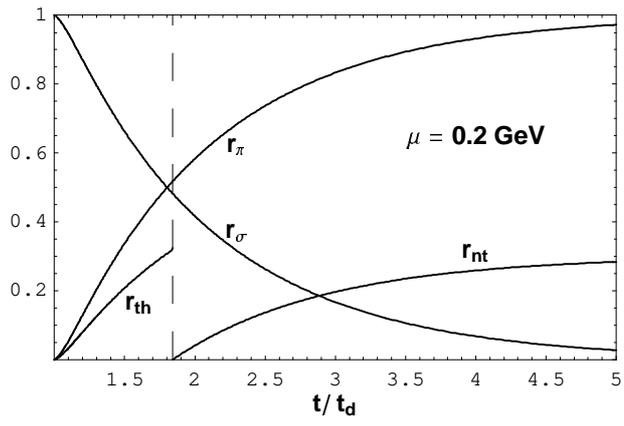
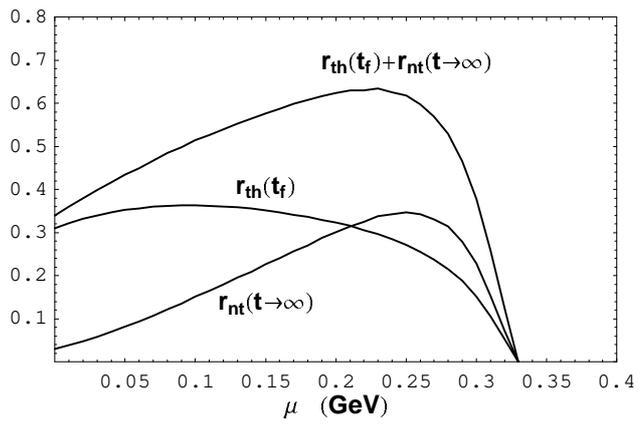
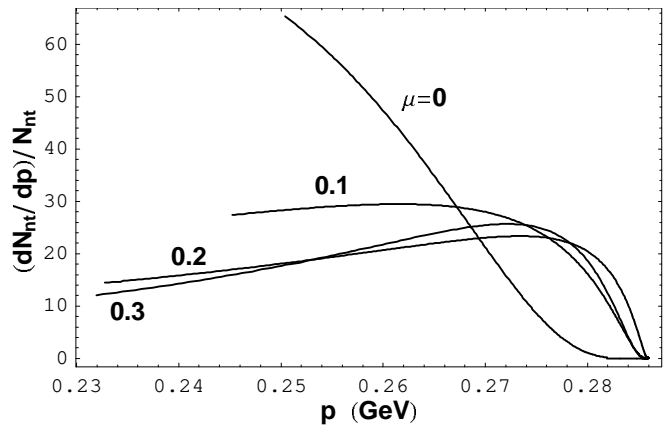
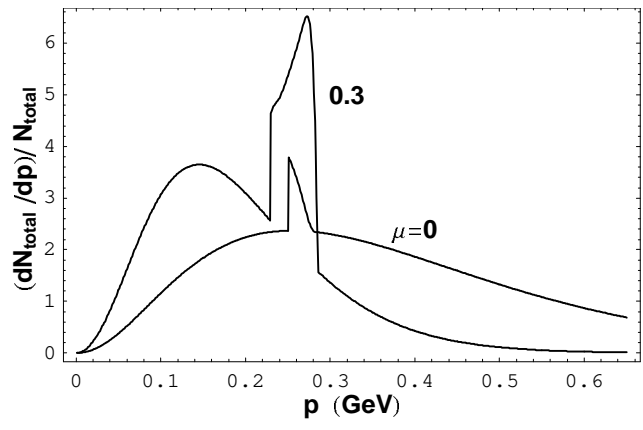
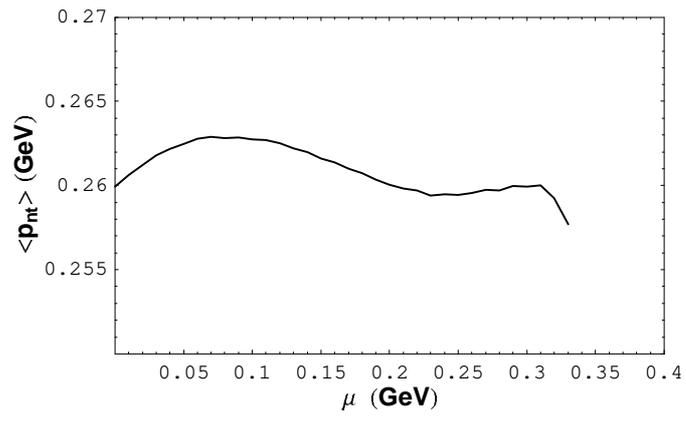
\end{document}